\documentclass[11pt,a4paper]{article}  

\usepackage[a4paper, top=1in, left=1in, right=1in, bottom=1in]{geometry}

\date{}
\usepackage{braket}
\usepackage{graphicx}
\usepackage{amsmath,amssymb,amsfonts}
\usepackage{hyperref}
\usepackage{bbold}
\usepackage{caption}
\usepackage{amsmath}  
\usepackage{amssymb}   
\usepackage{braket}
\usepackage{amsthm}  

\usepackage{changepage}

\usepackage{changepage}
\usepackage{breqn}
\usepackage{amsmath}
\usepackage{amsfonts}
\usepackage{tikz}
\usepackage[mathscr]{euscript}
\usepackage{yfonts}
\usepackage{tikz-cd}

\usetikzlibrary{positioning}

\usepackage{graphicx}
\usepackage[utf8]{inputenc}
\usepackage{amssymb}
\usepackage{amsthm}  
\usepackage{subcaption} 
\usepackage{url}
\usepackage{tikz}

\title{\large Effect of ambient gas on cavity formation for sphere impacts on liquids}

\begin{document}
\maketitle
\date{}
Hollis Williams$^1$, James Sprittles$^2$, Juan C. Padrino$^3$, and Petr Denissenko$^1$

$^1$School of Engineering, University of Warwick, Coventry CV4 7AL, UK

$^2$Mathematics Institute, University of Warwick, Coventry, CV4 7AL, UK 

$^3$School of Engineering, Newcastle University, Newcastle upon Tyne, NE1 7RU, UK
\newline
\newline
Contact email: $^1$Hollis.Williams@warwick.ac.uk, P.Denissenko@warwick.ac.uk $\\$ $^2$J.E.Sprittles@warwick.ac.uk, $^3$padr0006@umn.edu
\newline

	\begin{abstract}\noindent
Formation of a splash crown and a cavity following the impact of a sphere on a body of liquid is a classical problem.  In the related problem of a droplet splashing on a flat surface, it has been established that the properties of the surrounding gas can influence the splashing threshold.  At lower impact speeds, this is due mainly to the influence of gas kinetic effects, since the height of the gas lubrication film which is displaced during dynamic wetting is often comparable to the mean free path of the gas.  At higher Weber and Reynolds numbers, on the other hand, inertial effects dominate and the density of the gas becomes important in determining whether a splash occurs.  In this article, sphere impacts on a liquid body are investigated in a rarefied atmosphere using high-speed photography. It is found that the threshold entry speed for cavity formation is influenced by the density of the surrounding gas, whereas changing the mean free path of the gas has no effect.  We attribute this phenomenon to the gas slowing the sealing of the thin crown sheet behind the sphere.  This assertion is supported with experimental measurements of the liquid sheet thickness.  In the range of parameters considered, the splash crown influences the movement of the contact line, an effect not previously observed.


	\end{abstract}

\section{Introduction}

 A sphere plunging through the free surface of a body of liquid may splash and form a cavity which trails behind it. Air-water entry of projectiles has been studied since the nineteenth century \cite{Worthington1897,Worthington1908}.  Early studies, based mostly on the work of von Kármán and Wagner \cite{karman1929, wagner}, assumed that one can neglect the effects of liquid viscosity and surface tension due to large Reynolds and Weber numbers.  These studies also neglect the presence of surrounding gas due to its low density.  Subsequent work has shown that this picture is oversimplified and that the ambient gas influences dynamics of the cavity following a liquid-solid impact \cite{Gilbarg1948, May1952, Yakimov1975}.  Furthermore, Duez et al. established that both viscosity and surface tension of the liquid play a role in cavity formation, finding also that wettability of the impacting body is an important factor in determining the threshold speed $U^*$ for air entrainment \cite{duez2007}. Later, Marston et al. found that a falling sphere entraps a small amount of air at the bottom of the sphere due to the air pocket which forms because of the lubrication pressure in the gas layer between the sphere and the liquid surface.  This phenomenon is due to the air pocket which forms because of the lubrication pressure in the gas layer between the sphere and the liquid surface prior to impact.  As the liquid surface deforms, a thin sheet of air is produced which contracts to a bubble at the south pole of the sphere \cite{marston2011}.  Numerical simulations have found that cavity formation is connected with pinning of the contact line around the sphere and that viscosity of the gas in the film between the sphere and the liquid plays a role during early impact \cite{ding2015, yang2019}.  

There have been many studies of other variants of the falling sphere problem, although none have focussed directly on the influence of the surrounding gas on cavity formation \cite{dermeer2016}.  Aristoff et al. considered buoyant low-density spheres, finding expressions for the pinch-off time of a cavity and the volume of air entrained by the sphere \cite{aristoff2010}.  Hurd et al. studied the water entry characteristics of deformable elastomeric spheres, finding that the oscillations of these spheres during impact results in new types of nested cavities \cite{hurd2017}.  Watson et al. examined spheres with heterogeneous wetting properties, finding that spheres which are partly hydrophilic and partly hydrophobic always have asymmetric cavities and drift away from straight-line trajectories \cite{watson2021}.  Marston et al. observed cavity formation for heated sphere impacts, finding that there is an inverted Leidenfrost effect when the sphere temperature is much larger than the boiling point of the liquid, which either produces a cavity with smooth walls or a double cavity structure \cite{marston2012}.  Mansoor et al. also studied superhydrophobic spheres in detail and used a splash-guard mechanism to eliminate the phenomenon known as surface seal \cite{mansoor2014}.  Related studies have considered projectiles with varying aspect ratios and impacts on a two-phase fluid \cite{tan2014, smolka2019, kim2019}.  

The process of splash curtain formation in the sphere impact problem shares some similarities with splashing of a liquid droplet, but is believed to be driven by different physical mechanisms \cite{marston2015, marston2016a}.  Nevertheless, the similarities which exist might lead one to wonder whether the dynamics of the splash curtain can be influenced by the properties of the surrounding gas, given that it is now well-established that the properties of the surrounding gas play a significant role in influencing droplet splashing.  This line of investigation was initiated by Xu et al. who observed that splashing of a droplet when it impacts against a flat smooth surface can be completely suppressed by reducing the pressure of the surrounding gas \cite{xu2005}. Xu studied the dependence of droplet splashing on the roughness and texture of the surface, confirming that there is a different type of splashing caused by surface roughness (called prompt splashing) which must be distinguished from corona splashing on a smooth surface due to the presence of the surrounding air \cite{xu}. Benkreira and Khan demonstrated that air entrainment in the related problem of a dip coating flow can also be suppressed under reduced pressures and attributed this effect to the mean free path of the gas \cite{benkreira2008}.  

The regimes considered by these authors are typically for low impact speeds, where gas kinetic effects are important and the mean free path plays a significant role.  This occurs because the maximum speed at which the liquid can wet the solid surface is controlled by the speed at which the wetting gas lubrication film in front of the moving contact line is displaced.  The height of these films is typically extremely thin (of the same order as the mean free path of the gas) and as a consequence both droplet splashes and dip coating flows can be successfully described by models which incorporate kinetic effects \cite{sprittles, gordillo, gordillo2}.  As an example, the model of \cite{sprittles} uses kinetic theory in the gas film as described by the Boltzmann equation and combines this with regular hydrodynamics in the liquid phase as described by the Navier-Stokes equations.  

On the other hand, in the Gordillo-Riboux model for droplet splashing, the threshold speed for splashing is determined using the fact that splashing occurs due to a vertical lift force on the edge of the liquid sheet from the surrounding gas.  This lift force has two contributions: the lubrication force $\sim K_l \mu_g V_t$ and the suction force $\sim K_u \rho_g V_t^2 H_t$, where $\mu_g$ is the viscosity of the gas, $\rho_g$ is the density of the gas, $V_t$ is the initial velocity of the ejected lamella and $H_t$ is the initial height of the lamella.  $K_l$ and $K_u$ are coefficients which are derived from detailed calculations of the lift force in the region located between the lamella and the substrate (these calculations show that $K_l$ is approximately equal to a sum of two terms which both depend on the mean free path of the gas) \cite{gordillo}.  The lubrication force captures the viscous contributions to the lift force on the lamella edge and the suction force captures the inertial forces.  Note that viscosity is actually a mean free path effect, since changes in gas pressure do not change the viscosity of the gas.  During the characteristic impact time, viscous effects are confined to thin boundary layers with a typical width much lower than the radius of the droplet $R$.  Since the gas Reynolds number $\text{Re}_g$ based on $H_t$ and $V_t$ as the characteristic scales is $\sim \mathcal{O} (10)$, both the viscous and inertial contributions to the lift force must be considered.  The model suggests as expected that at higher Weber and Reynolds numbers the dominant forces in the droplet splashing problem for a smooth dry surface are inertial in nature \cite{gordillo}.  

Experimental evidence for this was found by Burzynski et al., who found that gas entrapmment is not the mechanism which is responsible for splashing at high Weber and Reynolds numbers and that splashing is influenced primarily by the density, not the mean free path, of the surrounding gas \cite{burzynski2019}.  This was done by considering a flywheel experiment and different gases at atmospheric pressure, with splashing outcomes analysed by measuring the size, velocity and angle of ejected secondary droplets.  The splashing outcome was also determined from the total volume ejected, where the theoretical expression for the volume is calculated using the Gordillo-Riboux theory \cite{gordillo}.  Guo et al. conducted numerical simulations of droplet impingement and splashing on dry and wet surfaces at very high impact speeds, finding that splashing on a dry surface can be suppressed by lowering the ambient gas density and that the properties of the ambient gas do not significantly influence splashing on a wet surface  \cite{guo2016}.  The simulations used a dynamic contact line model to define the boundary condition at the moving contact line and found that increasing the density of the gas increases the number of secondary droplets ejected from the lamella.   

The Gordillo-Riboux model has since been expanded by Pierzyna et al. using a data-driven threshold model which re-defines the threshold for droplet splashing on a dry smooth surface by collating a large number of experimental sources with different conditions and analysing the data with an uncertainty qualification analysis combined with machine learning \cite{pierzyna}.  The Gordillo-Riboux model can be considered as the special case where the parameter $\beta$ for determining splashing is a constant (in the general case, each impact parameter has a different dependence on $\beta$).  This more detailed threshold model incorporating a large number of diverse experimental results observed a linear dependence of the splashing threshold $\beta$ on impact speed $V$, surface tension $\sigma$ and gas density $\rho_g$ and an inversely proportional relation between $\beta$ and the liquid viscosity $\mu_l$.  This new model exhibits only weak dependency on the radius of the droplet $R$, liquid density $\rho_l$, gas viscosity $\mu_g$ and the mean free path of the gas $\lambda_g$.  These dependent variables are eliminated by a recursive feature elimination process to leave the four impact conditions mentioned above.  The eliminated variables are only irrelevant in their sense that they are irrelevant beyond the expression for $\beta$, so the Gordillo-Riboux captures all the influence of the eliminated variables on the splashing parameter, such that they are needed for any further corrections.

In this paper, we use high-speed photography to investigate the influence of the surrounding gas on cavity formation following the impact of a smooth dry wettable sphere on the free surface of a liquid body.  We begin in Section 2 by varying properties of both the solid and the liquid, finding agreement with existing experimental data on threshold speeds for cavity formation.  We then investigate the influence of the surrounding gas on cavity formation by using air, argon, and helium at different pressures hence varying gas density and the mean free path, obtaining results which diverge from the existing theoretical models.  In particular, we observe that cavity formation following a sphere impact on a liquid surface can be completely suppressed by reducing the pressure of the gas.  This is the first time that the transition from a cavity to a no cavity event has been effected only by changing the properties of the ambient gas.  We find that the most important parameter for influencing cavity formation is the density of the ambient gas and that the mean free path of the gas has negligible influence.  In Section 3, we provide a theoretical explanation for this effect based on the phenomenon by which the ambient gas influences the sealing of the crown sheet behind the sphere.  We finish in Section 4 with conclusions and suggestions for future work.

\section{Core Experimental Results}
 
 Spheres were released from an electromagnet above a container of silicone oil located inside a depressurised chamber.  The chamber was filled with air, argon, or helium at pressures from 0.05~bar to atmospheric.  Silicone oil was used instead of water to reduce the threshold entry speed for cavity formation and to prevent damage to the vacuum pump by water vapour. The vapour pressures of the oils used are all less than 0.002~bar, well below the pressures used in experiments, so that the influence of oil evaporation can be neglected.  

\begin{figure}
\includegraphics[width=16cm]{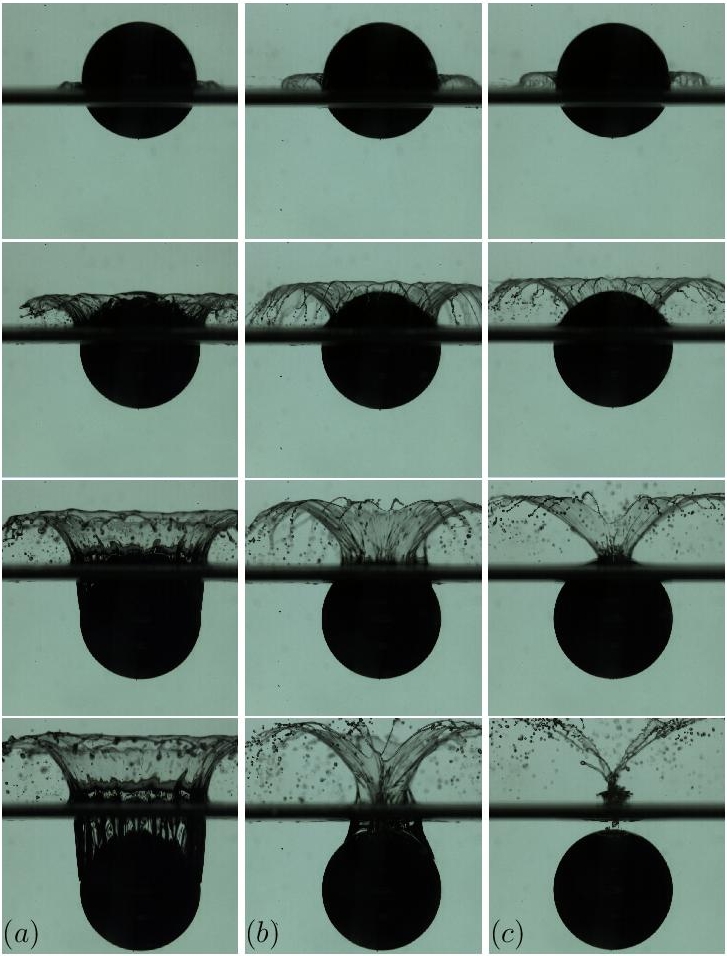}
\caption{Chronophotography of a 15 mm sphere plunging into 1.74~cP silicone oil at $V = 2$~ms$^{-1}$ at times $1.6$, $3.2$, $4.8$, and $6.4$~ms after touchdown.  The pressures are (a) $1$ bar, (b) $0.53$ bar, and (c) $0.5$ bar, respectively.  Videos \#1, \#2, and \#3 in the supplementary material}. 
\label{fig:splashchrono}
\end{figure}

The spheres were made of untreated chrome steel with density of $7.8$ g/cm$^3$ and static contact angle $\theta \approx 20-25^{\circ}$ which makes them oleophilic. The mean square roughness of the spheres was measured using a Bruker Contour profilometer at $0.3\,\mu$m, well below the thickness of the viscous boundary layer, reaching $\mathcal{O} (100\,\mu \text{m} )$ at the sphere equator.  Typical image sequences acquired when a sphere of diameter 15~mm plunges into the oil at atmospheric pressure, at 0.53~bar, and at 0.5~bar are shown in Fig.~\ref{fig:splashchrono}, demonstrating that reduced pressures can suppress cavity formation.  The impact of the sphere into the liquid was captured against an LED backlight with a Photron FASTCAM SA-X2 high-speed camera at a frame rate of 12 500 fps. To provide consistent surface wettability, the sphere was washed after each experiment with dilute ethanol, and then washed twice with deionised water.

\begin{figure}[!h]
\centering
\includegraphics[width=\columnwidth]{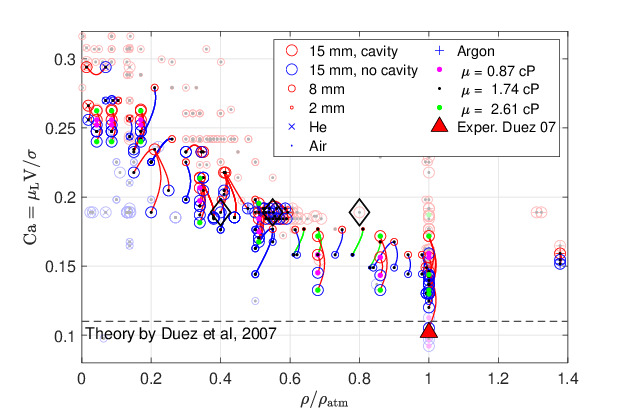}
\caption{Diagram of cavity/no cavity regimes at varying gas density ratio $\rho/ \rho_{\text{atm}}$ and capillary number $\text{Ca}$.  Three parameters are varied as indicated in the legend: sphere diameter, molecular mass of the gas, and dynamic viscosity of the liquid. Cavity/no cavity pairs with the same sphere size, oil viscosity, and gas type are connected by a line to indicate uncertainty of the cavity/no cavity threshold (red, blue, and green lines are used for the 15, 8, and 2~mm balls, respectively).  Black diamonds indicate regimes shown in Fig.~\ref{fig:splashchrono}. The red triangle and the black dashed line correspond to the experimental data and the theoretical prediction from \cite{duez2007}.} 
\label{fig:splashdata}
\end{figure}

A series of experiments was conducted using spheres with diameters $2$, $8$, and $15$ mm; silicone oils with viscosities $\mu_L = 0.87$, $1.74$, and $2.61$~cP; and entry speeds $V$ from 1 to 3.5~ms$^{-1}$.   The oil density varies little with viscosity, being $\rho = 0.9$, $0.87$, and $0.9$ g cm$^{-3}$, respectively, and the surface tensions are measured to be in the range $\sigma$ between $15$ and $20$~mN/m.  Three different gases are used for the surrounding atmosphere (helium, air and argon) which have molecular masses $M = 4$, $29$, and $39$ Daltons, respectively.  Since the mean free path of the gas $\lambda_g$ depends on the molecular mass, this allows us to separately isolate the effects of changing the density and the mean free path of the gas.  Results are shown as points in Fig.~\ref{fig:splashdata} in the parameter space of the capillary number $\text{Ca} = \mu_{\rm L} V / \sigma$ and the normalised gas density $\rho / \rho_{\text{atm}}$. No-cavity events are plotted with blue markers and the impacts where a cavity was formed are shown in red.

\begin{figure}[!h]
\centering
\includegraphics[width=120mm]{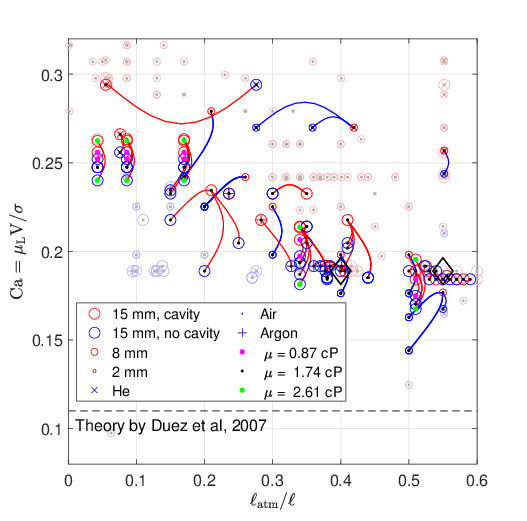}
\caption{Diagram of cavity/no cavity regimes similar to that in Fig.~\ref{fig:splashdata} with the inverse mean free path $\ell_{\text{atm}} / \ell$ at the horizontal axis. Note that, although a trend related to the change in gas density is present, a number of experiments with He prevents us from plotting an unequivocal divider between splash and no splash regimes.} 
\label{fig:splashdata_mfp}
\end{figure}

Our main qualitative result is that, in a range of parameters, the gas density influences whether a cavity forms during an impact on the free surface of a liquid. The molecular mass of the gas and the pressure seem to have no effect provided the density is kept constant, as confirmed by Fig.~\ref{fig:splashdata_mfp} where the mean free path is plotted at the horizontal axis, which is in stark contrast to experiments on droplet splashing \cite{xu2005}.  We note that the speed plateaus in a region of large gas densities exceeding 0.6 atmospheric density. The short range of densities over which the threshold speed varies suggests that the curve should instead be considered as a transition between two regimes: one where the role of the ambient gas is important and one where the influence of the gas is negligible.  In all cases, the transition between cavity and no-cavity regimes occurs in a threshold manner and is prone to perturbations in experimental conditions such as rotations of the falling sphere. This is confirmed by occasional observations of no cavity cases at unusually high entry speed such at the highest no-cavity point in Fig.~\ref{fig:splashdata}. At gas pressures below 0.05 bar, when the dynamic pressure of the liquid becomes close to the gas pressure, the oil boils upon impact and hence this range is not considered.

\section{Experimental Observations}
\noindent
Similarly to our results, the approximate independence of threshold speed $U^*$ on the sphere size for wettable spheres was previously reported by \cite{speirs2019} over a similar range of diameters and the inversely proportional dependence of threshold velocity on liquid viscosity was reported in \cite{duez2007}, i.e. for a given gas the transition is characterised by the capillary number. In contrast, the dependence of the splashing regime on the density of the surrounding gas observed in Fig.~\ref{fig:splashdata} contradicts the experimental results of Duez et al. \cite{duez2007}, who \emph{verified that the diameter of the impacting sphere does not influence the threshold, nor does the gas pressure (varied between 0.1 and 1 atm)}, a range of pressures in which we observe an approximate doubling of the critical speed.


From our experiments, we distinguish three main scenarios for sphere impact:

{\bf (a)} Flow-detached cavity: the flow detaches from the sphere around the equator, forming a vertically growing crown as shown in Fig.~\ref{fig:splashchrono}a;

{\bf (b)} Flow-attached cavity: the flow remains attached, i.e. wetting the sphere, forming a crown which starts closing behind the sphere but whose contact line does not reach the axis near the ball surface due to being slowed down by the gas (Fig.~\ref{fig:splashchrono}b);

{\bf (c)} No cavity: the flow remains attached, forming a crown whose base closes at the axis behind the ball before it is fully submerged below the unperturbed liquid surface (Fig.~\ref{fig:splashchrono}c).

\begin{figure}
   \centering
    \includegraphics[width=\columnwidth]{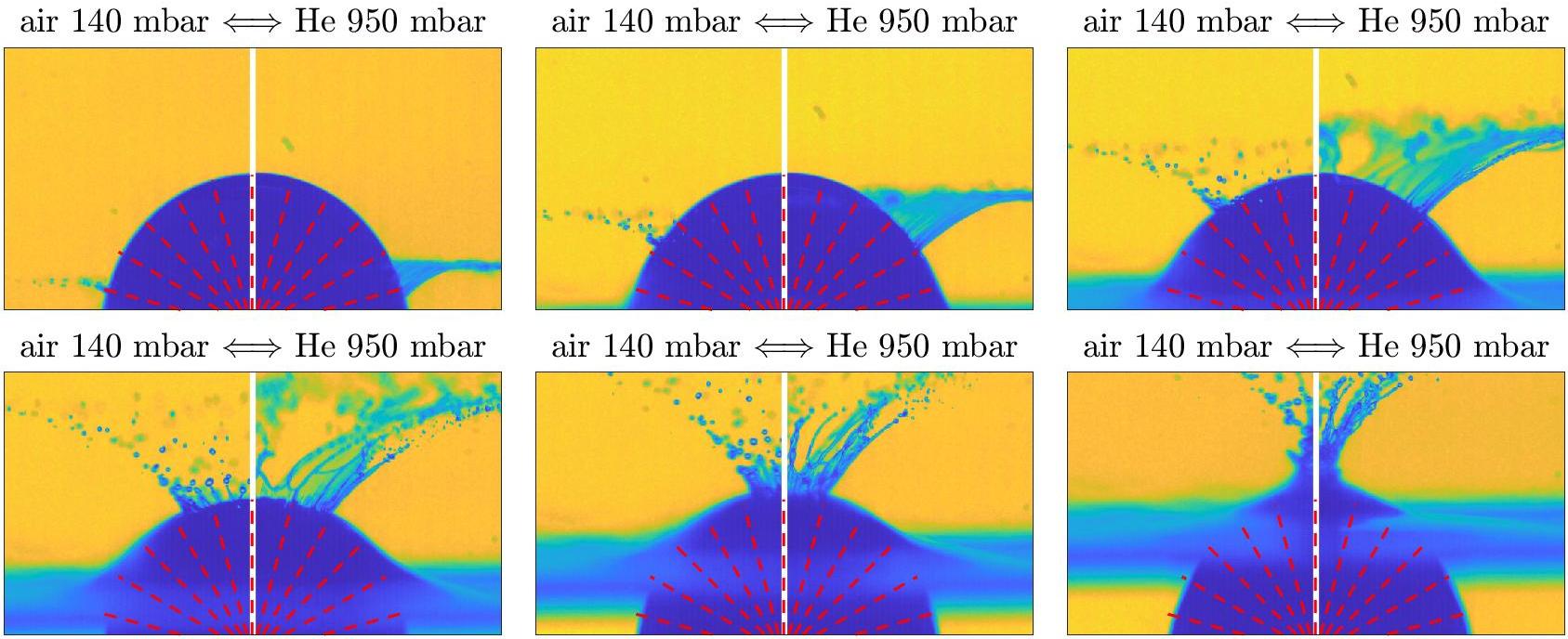}
        \caption{Testing the influence of gas molecular mass (and thus mean free path). Comparison of the impact in air at $p=140$~mbar (left panels of each image) and in helium at $p=950$~mbar (right panels).  Both cases correspond to the same gas density of 0.14~atmospheric.  The sphere has diameter 15 mm and impact speed $V = 2.06$ ms$^{-1}$.  Time separation between frames is 1~ms.  Videos \#4 and \#5 in the supplementary material.}
        \label{fig:compare_gases}
\end{figure}

\begin{figure}
    \centering
    \includegraphics[width=\columnwidth]{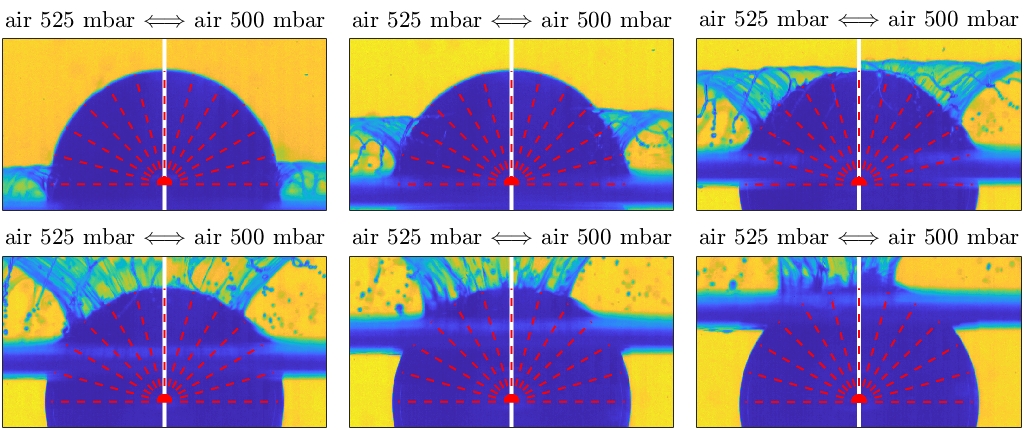}
       \caption{Testing the influence of gas density. Comparison of the impact in air at the density of 0.525~atmospheric (left panels) and in air at the density of 0.5~atmospheric (right panels). The sphere has diameter 15 mm and impact speed $V = 2.06$ ms$^{-1}$. Time separation between frames is 1~ms. Parameters are chosen close to the threshold so that the increase in the gas density by 5\% leads to cavity formation.  Videos \#2 and \#3 in the supplementary material.}
       \label{fig:compare_rho}.   
\end{figure}

To understand the effect of varying the ambient pressure, compare snapshots of a no-cavity event at the same gas density but with two different gases, air and helium, shown in Fig.~\ref{fig:compare_gases} in the sphere's frame of reference. Observe that the speed with which the liquid sheet rises up around the sphere is similar i.e. molecular weight / mean free path appears to have no effect on crown formation (although appearance of the crown and the ejection of droplets differs). Next consider time sequences at higher air densities, one slightly below the cavity/no-cavity threshold, and one slightly above, shown in Fig.~\ref{fig:compare_rho}. By comparing the right-hand side of Fig.~\ref{fig:compare_rho} with Fig.~\ref{fig:compare_gases}, observe that the speed of the crown propagation towards the pole is lower at higher gas density. When the gas density becomes too high (left-hand panels of Fig.~\ref{fig:compare_rho}), polar propagation of the crown slows down far enough for the contact line to not join at the apex and it instead goes below the liquid surface so that the cavity forms.


We now proceed to sketch the dynamics of the splash crown shown in Fig.~\ref{fig:splashincrownschematic} and explain how it is altered by the ambient gas. The influence of liquid evaporation on droplet splashing at reduced pressure was previously ruled out \cite{sefiane, xucomment}.  Similarly, liquid evaporation only influences contact line motion for sufficiently slow processes involving volatile liquids, so can be discounted \cite{rednikov}. The fact that the threshold speed only depends on the gas density suggests that the effect is related to the dynamic pressure of the gas, i.e. to inertial effects. Furthermore, independence of the threshold speed on the sphere radius suggests that the effect might relate to deceleration of the crown sheet itself which, in turn, slows down propagation of the contact line.

\begin{figure}
    \centering
    \includegraphics[width=150mm]{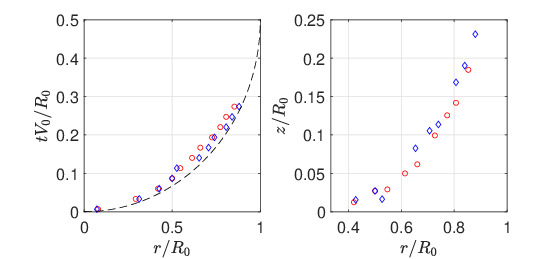}
        \caption{Growth rate for wetted part of sphere against time.  The dashed black line shows the radius of the cross-section of the sphere at the level of the liquid. Red markers correspond to the experiment where cavity was formed (air at 1000~mbar), blue markers to the experiment with no cavity (air at 150~mbar).  The impact speed is $V \approx 2.3$ ms$^{-1}$  in both cases. } 
        \label{fig:compare_wettedcircle}
\end{figure}

Figure~\ref{fig:splashchrono}  suggests that the difference between cavity/no-cavity appears above the unperturbed liquid level and above the equator of the sphere.  This was confirmed by filming the experiment from beneath the free surface to measure the growth rate of the wetted circle below the equator, similar to experiments by \cite{kim2019}.  As seen in Fig.~\ref{fig:compare_wettedcircle}, evolution of the wetted region of the sphere with and without cavity formation is similar at the lower hemisphere, hence we focus on the upper hemisphere and the stage when the crown appears.  The black dashed line indicates the free-fall trajectory which would be expected for the displaced liquid from geometric considerations, indicating that there is no influence on the motion of the contact line from trapped air in the pre-impact layer between the solid and the liquid and that cavity formation is not due to gas entrapment.  We therefore eliminate the possibility that the observed shift of the threshold entry speed is related to the gas trapped in the layer between solid and liquid and instead argue that the difference is due to the modification of the dynamics of the splash crown above the equator by the surrounding gas.

The advance of the root of the crown (and hence the contact line) is where the gas plays its role, separating the cavity and no-cavity regimes. By comparison of Fig.~\ref{fig:compare_rho} and Fig.~\ref{fig:compare_gases}, we observe that the liquid-solid angle at the root of the crown is close to 90$^\circ$ at low gas density and decreases at higher gas densities. Here, one should distinguish between the advancing (apparent) contact angle at the liquid-solid boundary and the angle which the crown sheet makes with the surface of the sphere, which is defined by the ratio between the speed of the crown root and the polar velocity of liquid feeding into the crown.  It follows from the equality of the arc length and the tangent length for sufficiently small arc angles that when the speed of the crown root and the polar velocity are the same, the angle is 90$^\circ$.  If this angle falls below $90^\circ$, the speed of the advancing contact line decreases.  Near the cavity/no cavity threshold, the ambient gas slows the speed of the contact line by up to 50\%, resulting in a 50\% decrease in the threshold entry speed as seen in Fig.~\ref{fig:splashdata}. 

\begin{figure}
    \centering
    \includegraphics[width=120mm]{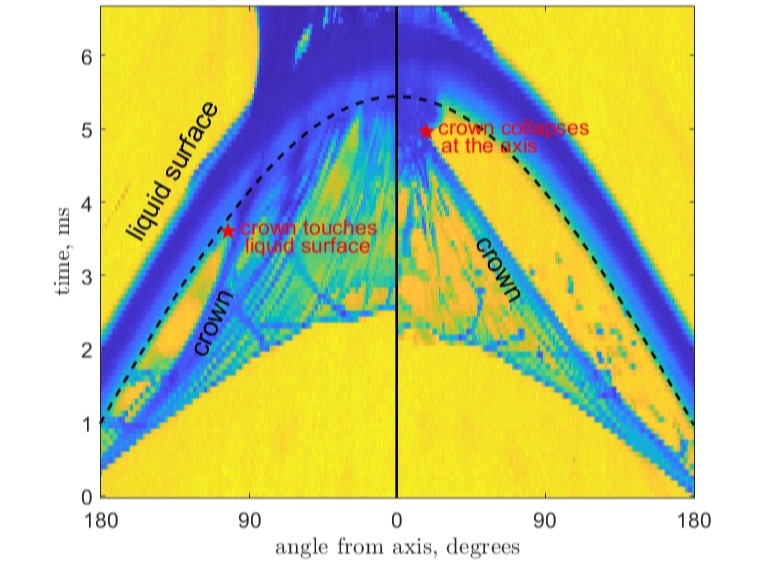}
       \caption{Spatiotemporal image of the curtain propagating towards the pole in the experiments shown in Fig.~\ref{fig:compare_rho} of the main text (left panel is air at $p=525$ mbar and right panel is air at $p=500$ mbar). The dashed line indicates position of the unperturbed liquid surface. At lower gas density (right panel), the crown deceleration is not sufficient to prevent it from collapsing at the pole. At higher gas density (left panel), deceleration is higher and the liquid surface overtakes the crown, forming a cavity.}
       \label{fig:spatiotemporal}
\end{figure}

Slowing down of the crown is shown in detail in Fig.~\ref{fig:spatiotemporal}, where we plot the spatiotemporal image illustrating propagation of the crown towards the pole. The image was created by stacking the semi-circular strips adjacent to the sphere at sequential time instants. The dashed sinusoidal arch represents advance of the horizontal liquid surface along the sphere and its shape is dictated by basic geometric considerations. The root of the crown is seen to decelerate and is either overtaken by the liquid surface or collapses at the pole.  In the first case, the cavity forms and in the latter case, the liquid surface closes with no cavity as a result.

\begin{figure}
    \centering
    \includegraphics[width=140mm]{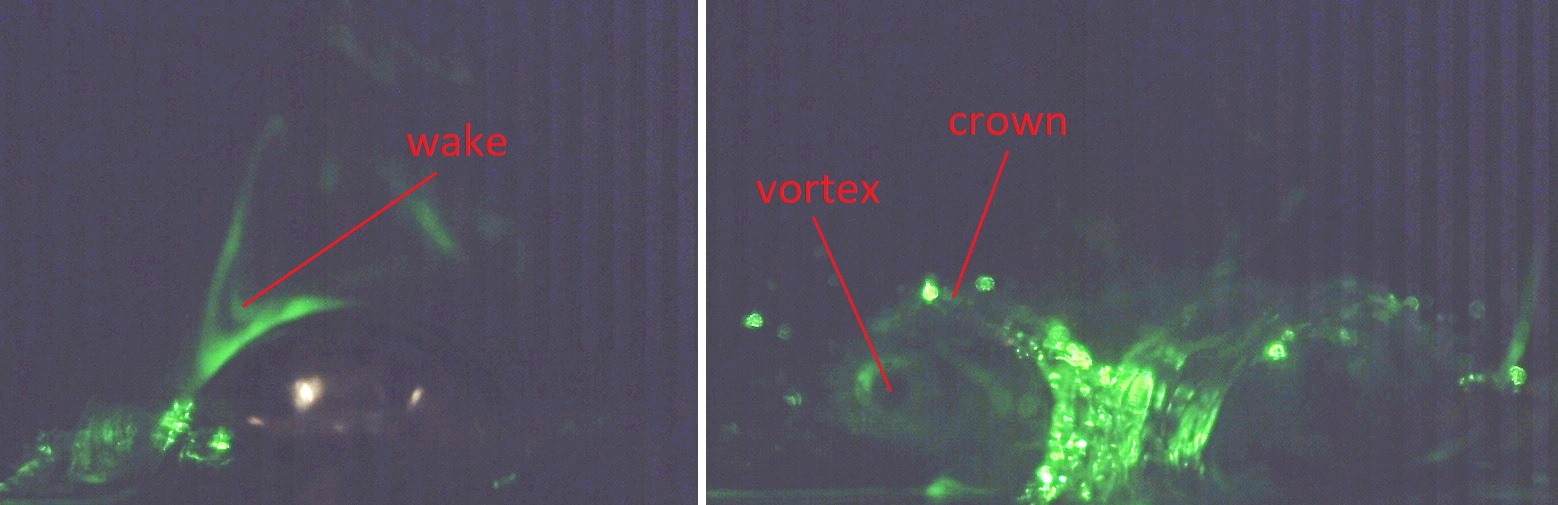}
       \caption{Oil vapour and lasersheet visualisation of the gas flow behind the falling sphere (left) and of the vortex formed behind the crown. $V \approx 1.6$ ms$^{-1}$, air at atmospheric pressure.   The bottom of the image coincides with the unperturbed liquid level.  Video \#8 in the supplementary videos.}
       \label{fig:lasersheet_visualisation} 
\end{figure}

A natural question one might ask is whether the gas flow around the sphere as it falls towards the liquid surface has an influence on the splash crown.  To check this, in Fig.~\ref{fig:lasersheet_visualisation} we show a laser sheet visualisation with smoke as the seed material and air at atmospheric pressure.  The resulting flow visualisation is as expected, with the presence of a wake behind the falling sphere which then flows out of the sealing crown and forms a pair of vortices either side, so we also rule this out as the reason for cavity formation.

\begin{figure}
    \centering
    \includegraphics[height=80mm]{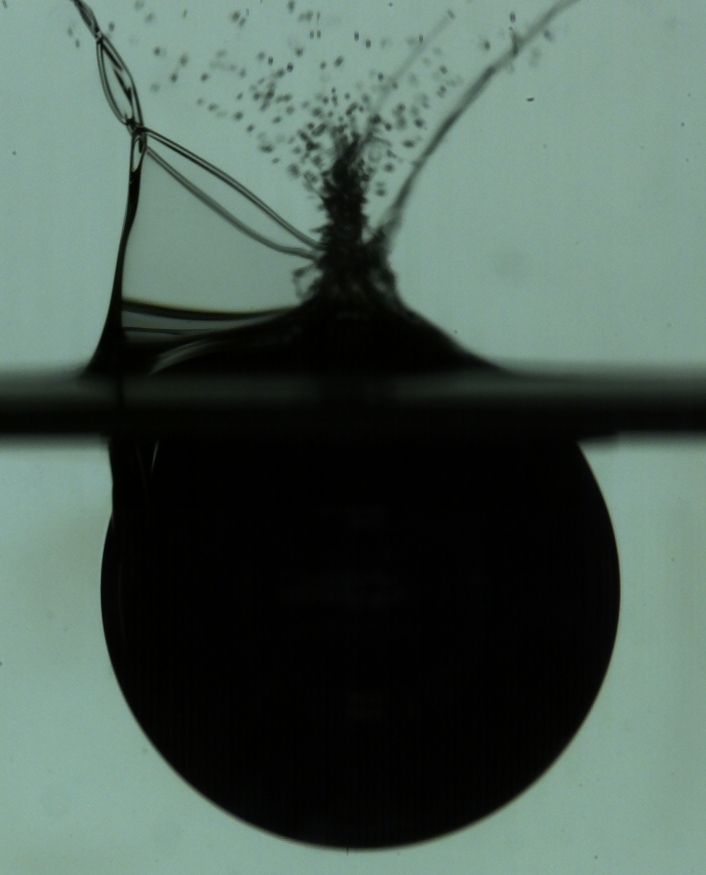}
    \includegraphics[height=80mm]{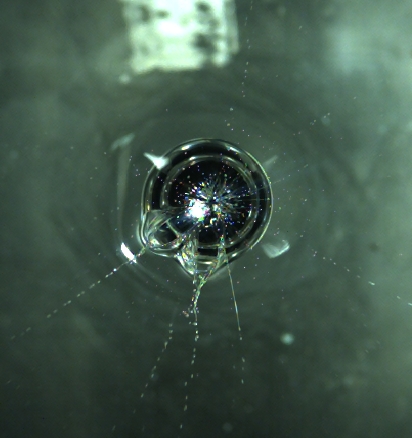}
       \caption{Asymmetric splash at low gas density at higher impact speed: Formation of cavities on one side while the most of circumference is cavity-free. $V \approx 2.8$ ms$^{-1}$, Helium, $p=138$~mbar left, side view and $p=200$~mbar right, top view.  Videos \#6 and  \#7 in the supplementary videos.} 
       \label{fig:asymmetric}
\end{figure}

We see from Fig.~\ref{fig:splashdata} that a cavity forms at sufficiently high impact speeds, even when the gas density is very low.  We suggest that this plateau at low gas densities is due to the fact that in this regime the surrounding gas does not slow the crown and that there are instead instabilities or rotational effects such that a cavity almost always forms for sufficiently fast impacts on the liquid surface.  In Fig.~\ref{fig:asymmetric}, the case where the crown collapses at the pole unimpeded is shown.  However, the contact line propagation becomes unstable and the flow separates at some point, creating an asymmetric cavity.

\begin{figure}
    \centering
    \includegraphics[width=160mm]{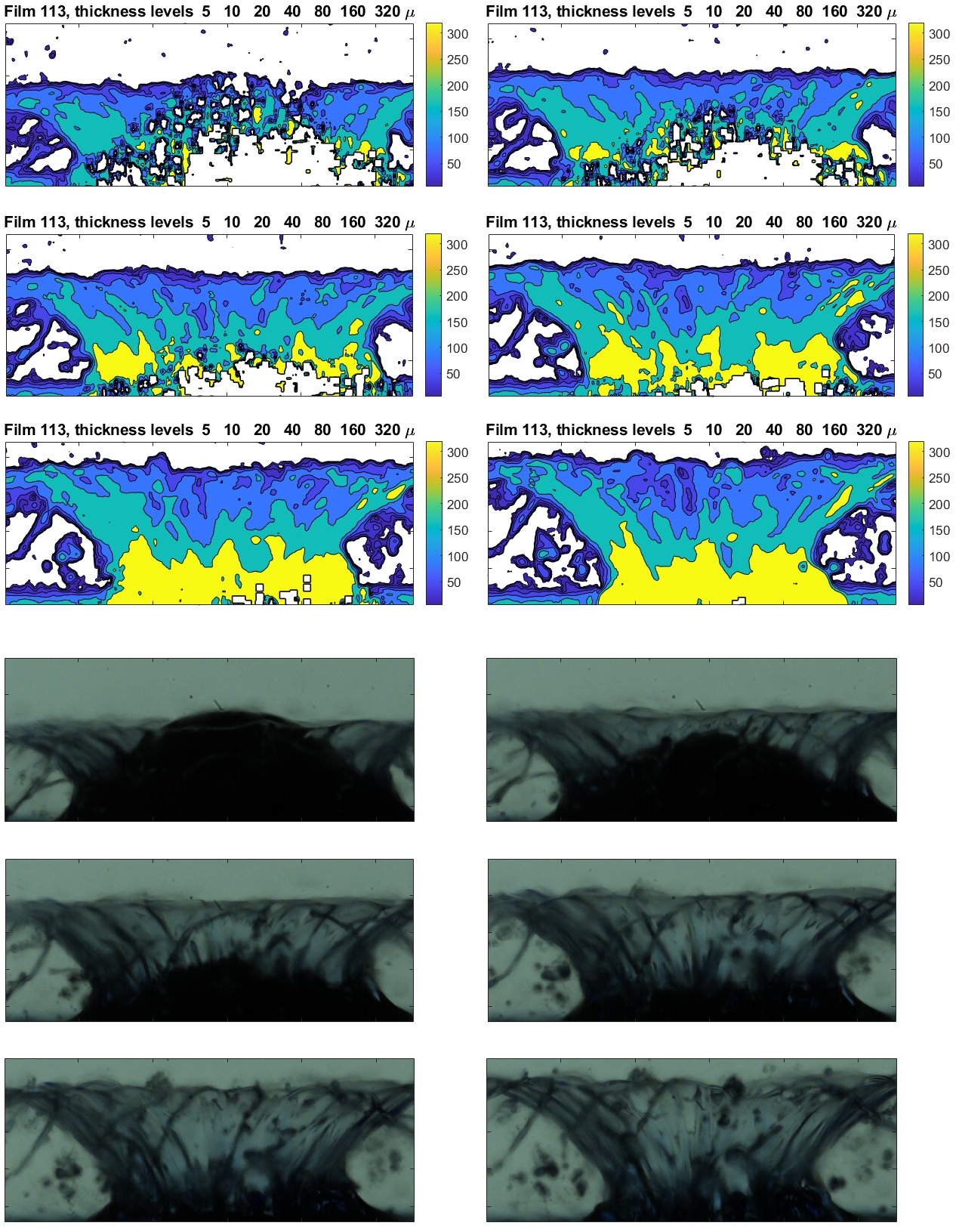}
    \caption{Sheet thickness measured by the blue-coloured oil: Contour levels of the liquid sheet thickness and the corresponding optical images. Atmospheric pressure, 13~cm drop height. Levels are plotted with the 2-fold increments in the liquid thickness. Observe the thickness corresponding to the major part of the curtain corresponds to the interval 80-160~micron so that, accounting for the two parts of the curtain captured, the thickness of the curtain is half this value, approximately 50-100~$\mu$m.} 
        \label{fig:thickness_13cm}
\end{figure}

In order to provide understanding and aid the development of a model, the thickness of the liquid crown was measured by adding blue dye to the oil. The thickness of the liquid sheet is measured by the ratio of red channel in the colour image to the total brightness. Levels are calibrated using a wedge-shaped cuvette with known thickness. In Fig.~\ref{fig:thickness_13cm}, visible images of the splash are shown together with the corresponding thickness maps. Taking into account the fact that at each location we observe two layers of the crown, its thickness can be evaluated at the level of 50-100~$\mu$m.

\section{Discussion}
\noindent
Summarising the above experimental observations, we propose the following sequence of events on impact:

{\bf (1)} The sphere displaces liquid which, rising along the sphere, forms a collar. Viscous effects are negligible in the formation of the bulk of the collar as the liquid boundary layer thickness $\delta\sim\sqrt{\mu/\rho_{\rm liq}\cdot R/V_0}\sim100~{\rm \mu \text{m}}$, whilst the thickness of the bulk of the collar is of the order of millimetres as seen in Fig.~\ref{fig:compare_gases}. Here, we assume that the velocity of the liquid in the collar is of the same order of magnitude as the impact velocity.

{\bf (2a)} Due to surface tension, a rim appears at the leading edge of the collar, similar to that observed for vertical cylinders \cite{kim2019}. At too high an impact speed, the flow separates from the sphere at the equator and the cavity forms without the crown propagating along the surface of the sphere as in Fig.~\ref{fig:splashchrono}a.

\begin{figure}
    \centering
    \includegraphics[width=70mm]{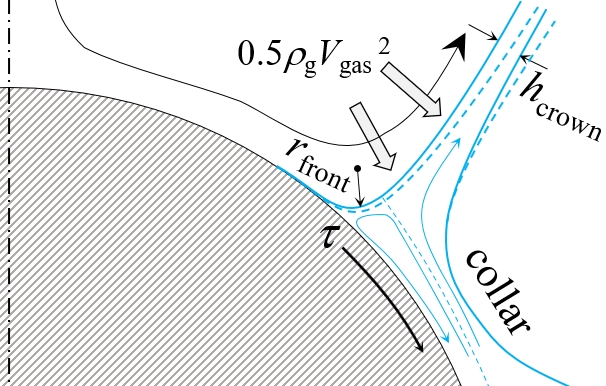} 
    \caption{Schematic in the frame of reference moving with the root of the crown. In the absence of the ambient gas, propagation of the front is defined by the balance between capillary pressure under the concave liquid surface and viscous forces. When the ambient gas density is high enough, it slows the crown (dashed line), which changes the shape of the curved interface and slows down propagation of the liquid front.} 
        \label{fig:splashincrownschematic}
\end{figure}

{\bf (2b)} At lower velocities, the rim moves past the equator but starts to depart the surface due to centrifugal force, and the crown forms as shown in Fig.~\ref{fig:splashincrownschematic}. In the absence of gas and surface curvature, the velocity of propagation of the wetting front $V_{\rm front}$ would be defined by the balance between the pressure created by the concave surface formed due to the sphere wettability and the viscous stress in the vicinity of the front \cite{yang2019, xu2005}:
\begin{equation}
   \frac{\sigma}{r_{\rm front}} \sim \frac{\mu V_{\rm front}}{r_{\rm front}} \Longrightarrow V_{\rm front}\sim\frac\sigma\mu
   \label{eqn_Vfront_vac}
\end{equation}
which corresponds to $\text{Ca}_{\text{front}}\sim1$ as reported in \cite{duez2007}.  In our case with 1.74~cP oil, $V_{\rm front}\sim10$~ms$^{-1}$ which agrees with the speed measured from videos and also agrees with the observed threshold impact velocity of 2-3~m/s independent of the sphere radius. 

{\bf (2c)} Below the transition speed, the crown crown closes at the pole before the sphere is submerged and no cavity is observed. 

To confirm the plausibility of relating the cavity formation to deceleration of the crown by the surrounding gas, we present an order of magnitude estimate of the stresses involved in the problem at $V_0=2$ $\text{ms}^{-1}$ (drop height of 20 cm) and $D=15$~mm. The dominant term is the ambient pressure $P_0 > 5000$~Pa. This pressure provides continuity of the liquid so that the liquid displaced by the sphere moves up along the solid surface.  The next term is dynamic pressure of the liquid at $P_{\rm liq}\sim \rho_{\rm liq} V_0^2/2 \sim 2000$~Pa. Besides this, we have dynamic pressure of the gas $P_{\rm dyn\,gas}$. To evaluate this term, we suppose the crown speed relative to the gas $V_{\rm gas}$ to be given by the speed of the crown of $2V_0$ estimated from the experiment in Fig.~\ref{fig:compare_rho} and add another factor of $V_0$ for the circulatory flow which forms behind the falling sphere to find that the dynamic gas pressure at which the cavity/no-cavity transition occurs is given by

\begin{equation}
P_{\rm dyn\,gas}\sim \rho_{\rm gas} V_{\rm gas}^2/2 \sim \rho_{\rm gas} (3V_0)^2/2 \sim 10~\text{Pa}
\label{eqn_p_dyn_gas}
\end{equation}
The radius of curvature of the advancing front $r_{\rm front}$ shown in  Fig.~\ref{fig:splashincrownschematic} is defined by the balance of capillary force and centrifugal terms which accelerate fluid into the crown: 
\begin{equation}
    \frac{\sigma}{r_{\rm front}} \sim r_{\rm front}\rho_{\rm liq}\frac{V_0^2}{R} \Longrightarrow r_{\rm front}\sim\sqrt{\frac{\sigma R}{\rho_{\rm liq} V_0^2}}.
    \label{eqn_rfront}
\end{equation}
Reassuringly, for sphere diameters from 2 to 15 mm, this estimate for $r_{\rm front}$ varies from 70 to 200~micron, which does not contradict experimental observations. The corresponding capillary pressure varies from 150 to 400~Pa. The thickness of the crown $h_{\rm crown}$ is measured experimentally (Fig.~\ref{fig:thickness_13cm}), resulting in a radius-independent value of 50-100~micron for the 16~mm, 8~mm, and 4~mm spheres.

The deceleration $a_{\rm stop}$ required for the crown moving at the speed of $2V_0$ to completely stop at the pole, passing the distance of approximately $R$, is
\begin{equation}
    a_{\rm stop} \sim \frac{2V_0^2}{R} .
    \label{eqn_a_characteristic}
\end{equation}
The deceleration of the crown due to the dynamic pressure of the gas is
\begin{equation}
    a_{\rm crown} \sim\frac{P_{\rm dyn\,gas}}{\rho_{\rm liq}\,h_{\rm crown}}
    \label{eqn_a_crown}.
\end{equation}
The exact shape of the liquid surface adjacent to the advancing wetting front depends on the angle which the root of the crown makes with the surface of the sphere and affects the speed of propagation of the wetting front.  A sharper angle increases the curvature, increasing the capillary term in (\ref{eqn_Vfront_vac}) and hence the speed of the front, whereas angles above 90$^\circ$ lower the curvature and the speed of the front. In this way, by slowing the crown, the ambient gas slows the wetting front (\ref{eqn_Vfront_vac}), preventing it from reaching the pole before the falling sphere submerges below the surface. 
Combining (\ref{eqn_p_dyn_gas})-(\ref{eqn_a_characteristic}), we have
\begin{equation}
    \frac{a_{\rm crown}}{a_{\rm stop}} \sim \frac{\rho_{\rm gas} V_{\rm gas}^2}{2\rho_{\rm liq}h_{\rm crown}}\frac{R}{2V_0^2}
    \sim 2\,\frac{\rho_{\rm gas}}{\rho_{\rm liq}}\frac{R}{h_{\rm crown}}
    \label{eqn_a_ratio}
\end{equation}
 At the set of parameters when the cavity/no-cavity transition occurs (Fig.~\ref{fig:compare_rho}), $a_{\rm crown} / a_{\rm stop}$ varies between $1$ and $2$, depending on the sphere radius $R$.  This supports our hypothesis that the presence of the ambient gas shifts the threshold speed for cavity formation, manifesting in the lowering of the threshold $\rm Ca$ between 0.2 and 0.6 atmospheric density in Fig.~\ref{fig:splashdata}.  We emphasise that (\ref{eqn_a_ratio}) is an order of magnitude estimate valid within the considered range of parameters and not a scaling.    \section{Conclusions}

We have investigated the transition between regimes in which a cavity forms or fails to form during sphere impacts on a free liquid surface. We have found that the threshold speed for cavity formation is significantly influenced by the ambient gas, where the gas density is the defining parameter and the mean free path in the gas has negligible influence. The diagram of cavity/no-cavity regimes (Fig.~\ref{fig:splashdata}) shows two limiting cases: one at low gas density (below 0.2 atmospheric air density) and one at high gas density (above 0.6 atmospheric). At low gas densities, cavity/no cavity transition occurs due to instability in the contact line propagation and appears not to be affected by the surrounding gas. As the gas density increases, the dynamic pressure of the gas slows the propagation of the crown towards the pole of the sphere, reducing the threshold speed.

In this work, we have focussed on smooth oleophilic spheres, but it would certainly be an interesting avenue for further work to study the influence which surface wettability/roughness can have on the effect which we have discovered \cite{zhao}.  As explained in Section 1, the extension of the results on the influence of the surrounding gas for droplet splashing on a flat smooth surface in \cite{xu2005} to droplet splashes on rough and textured surfaces is somewhat involved, so we would expect the extension of our work to rough and textured spheres to be similarly difficult.  Splashes on smooth surfaces experience a type of splashing which is caused by the presence of the atmospheric gas, whereas splashes on rough or textured surfaces, with the different dynamics of contact line propagation, the effect of the ambient gas may differ.  The effect of corona deceleration can be completely isolated by pulling a high vacuum \cite{xu}. 

Studies of the influence of gas density on the sphere impact problem can be developed in many directions. A recent study has found that the gas density is also the most relevant quantity influencing the ejection of the lamella during droplet splashing at high Re and We, so we anticipate that our result may stimulate some models which attempt to unify the splashing mechanism which is found in the droplet splash and sphere impact problems in these regimes \cite{burzynski2019, guo2016}.  This is, however, a complicated problem and there are many other factors which affect the sphere splashing mechanism (including instabilities of the flow and contact line) which will need to be studied further in experiments.    Further experimental work can also be conducted to investigate the exact shape of the liquid surface at the root of the crown \cite{burzynski2018}.  

Our results will also complement the development of multi-scale computational simulation of air-water entry of spheres since the existing models have focused only on the effect of the viscosity, rather than the density, of the gas \cite{yang2019}.  We also anticipate that our findings will stimulate theoretical models which modify the classical Wagner model for impacts with the presence of ambient gas.  The existing models of this type only consider pre-impact or early-impact air-water interactions where viscous forces in the gas deform the liquid, whereas we have demonstrated that the gas continues to play a role long after the pre-impact stage \cite{moore}.  In terms of applications, it is interesting that the splash of a projectile which falls and hits a liquid surface can be suppressed by altering the properties of the ambient gas above the liquid, implying that it is possible for a projectile to slip through a surface unobserved more easily in a rarefied atmosphere.  This complements the existing observation that sphere splashing can be suppressed by using a smooth rather than a rough sphere \cite{zhao}.

\section*{Acknowledgements}

\noindent
This research has been financially supported by EPSRC Grants No. EP/N016602/1, No. EP/S029966/1, No. EP/P031684/1, and No. EP/P020887/1.

\end{document}